\documentclass[A4,10pt]{iopart}
\usepackage{amsfonts}
\usepackage{geometry}
\usepackage{graphicx}

\begin{document}

\title{Equilibrium fluctuation theorems compatible with anomalous response}
\author{L. Velazquez and S. Curilef}

\begin{abstract}
Previously, we have derived a generalization of the canonical
fluctuation relation between heat capacity and energy fluctuations $
C=\beta^{2}\left\langle \delta U^{2}\right\rangle $, which is able
to describe the existence of macrostates with negative heat
capacities $C<0$. In this work, we extend our previous results for
an equilibrium situation with several control parameters to account
for the existence of states with anomalous values in other response
functions. Our analysis leads to the derivation of three different
equilibrium fluctuation theorems: the \textit{fundamental and the
complementary fluctuation theorems}, which represent the
generalization of two fluctuation identities already obtained in
previous works, and the \textit{associated fluctuation theorem}, a
result that has no counterpart in the framework of Boltzmann-Gibbs
distributions. These results are applied to study the anomalous
susceptibility of a ferromagnetic system, in particular, the case of
2D Ising model. \newline
\end{abstract}
\address{Departamento de F\'\i sica, Universidad Cat\'olica del Norte, Av. Angamos 0610, Antofagasta,
Chile.} \maketitle

\section{Introduction}

According to the well-known fluctuation relation:
\begin{equation}
C=k_{B}\beta ^{2}\left\langle \delta U^{2}\right\rangle  \label{can.cfr}
\end{equation}%
between the heat capacity $C$ and the energy fluctuations $\left\langle
\delta U^{2}\right\rangle $, the heat capacity should be \textit{nonnegative}%
. However, such a conclusion is only an illusion. Since the first
theoretical demonstration about the existence of macrostates with negative
heat capacities $C<0$ by Lyndel-Bell in the astrophysical context \cite%
{Lynden}, this anomaly has been observed in diverse systems \cite%
{moretto,Dagostino,gro na,Lyn2,pad,Lyn3,gro1,Dauxois}. The
fluctuation relation (\ref{can.cfr}) directly follows from the
consideration of the Gibbs' canonical ensemble:
\begin{equation}
dp_{c}\left( U\left\vert \beta \right. \right) =\frac{1}{Z\left( \beta
\right) }\exp \left( -\beta U\right) \Omega \left( U\right) dU,  \label{can}
\end{equation}%
which accounts for the equilibrium thermodynamic properties of a system in
thermal contact with a heat bath at constant temperature $T$ when other
thermodynamical variables like the system volume $V$ or a magnetic field $H$
are kept fixed, where $\beta =1/k_{B}T$. As already commented, macrostates
with $C<0$ can be observed within the thermodynamic description of a given
system, but they are \textit{unstable} under the external influence imposed
to the system within the canonical ensemble (\ref{can}).

Fluctuation relation (\ref{can.cfr}) admits the following
generalization \cite{Vel.ETFR,Vel.Thirring}:
\begin{equation}
C=k_{B}\beta ^{2}\left\langle \delta U^{2}\right\rangle +C\left\langle
\delta \beta ^{\omega }\delta U\right\rangle ,  \label{unc}
\end{equation}%
which considers an equilibrium situation where the environmental
inverse temperature $\beta ^{\omega }$ exhibits correlated
fluctuations with the total energy $U$ of the system under study as
a consequence of the underlying thermodynamic interaction. The
\textit{feedback effect} $\left\langle \delta \beta ^{\omega }\delta
U\right\rangle $ consideration allows to detect the presence of a
regime with negative heat capacities $C<0$ in the microcanonical
caloric curve $1/T\left( U\right) =\partial S\left( U\right)
/\partial U$. In fact, it asserts that macrostates with $C<0$ are
thermodynamically stable provided that the environmental influence
obeyed the inequality $\left\langle \delta \beta ^{\omega }\delta
U\right\rangle >1$.

The energy-temperature fluctuation relation (\ref{unc}) has
interesting connections with diverse questions within statistical
mechanics, such as the justification of a complementary relation
between energy and temperature
\cite{Vel.ETFR,Vel.Thirring,Vel.URSM}, the extension of canonical
Monte Carlo methods to allow the study macrostates with negative
heat capacities and to avoid the super-critical slowing down of
first-order phase transitions \cite{Vel.MC1,Vel.MC2}, as well as the
development of a geometric formulation for fluctuation theory based
on the existence of \textit{reparametrization dualities} \cite
{Vel.Geo}. However, Eq.(\ref{unc}) is applicable to those
equilibrium situations where there is only involved the conjugated
pair energy-temperature. Consequently, this result merely
constitutes a special case of certain equilibrium fluctuation
theorems compatible with the existence of anomalous response
functions \cite {Ison,Einarsson,Chomaz,Gulminelli,Lovett,Hugo},
whose derivation will be the main goal of the present work.

\section{Equilibrium fluctuation theorems}\label{demostracion}

\subsection{Notations and conventions}

From the standard perspective of statistical mechanics, a
system-environment equilibrium situation with several control
parameters is customarily described using the
\textit{Boltzmann-Gibbs distributions} \cite{landau,Reichl}:
\begin{equation}
dp_{BG}\left( \left. U,X\right\vert \beta ,Y\right) =\frac{1}{Z\left( \beta
,Y\right) }\exp \left[ -\beta \left( U+YX\right) \right] \Omega \left(
U,X\right) dUdX.  \label{BGD}
\end{equation}%
The quantities $X=\left( V,M,P,N_{i},\ldots \right) $ represent
other macroscopic observables acting in a given application
(generalized displacements) as the volume $V$, the magnetization $M$
and polarization $P$, the number of chemical species $N_{i}$, etc.;
with $ Y=\left( p,-H,-E,-\mu _{i},\ldots \right) $ being the
corresponding conjugated thermodynamic parameters (generalized
forces) as the external pressure $p$, magnetic and electric fields,
$H$ and $E$, the chemical potentials $ \mu _{i}$, etc.

The notation employed in thermodynamics always distinguishes energy
and temperature from the other thermodynamic quantities. Such a
distinction is clearly evident in thermodynamic relations as
$dQ=TdS=dU+YdX$. In this work, we shall also adopt the following
convention for the generalized displacements $\left( U,X\right)
\rightarrow I=\left( I^{1},I^{2},\ldots\right) $ and $\left( \beta
,Y\right) \rightarrow\beta=\left( \beta_{1},\beta_{2},\ldots\right)
$ for the generalized forces, which allows us to deal with a
\textit{symmetric and compact notation} in the thermodynamic
expressions. Thus, the previous example is equivalently expressed as
follows: $dS=\beta\left( dU+YdX\right) \rightarrow
dS=\beta_{1}dI^{1}+\beta_{2}dI^{2}+\ldots\equiv\beta_{i}dI^{i}$.
Besides, we shall assume the Einstein's summation convention, which
allows to rewrite the probabilistic weight of the Boltzmann-Gibbs
distribution (\ref{BGD}) as:
\begin{equation}
\omega_{BG}\left( \left. I\right\vert \beta\right) =\frac{1}{Z\left(
\beta\right) }\exp\left( -\beta_{i}I^{i}\right) .  \label{bg.w}
\end{equation}
For convenience, Boltzmann's constant $k_{B}$ is hereafter assumed
as the unity.

\subsection{General fluctuation theorems of a classical distribution function}

Let us start from a generic classical distribution function:
\begin{equation}\label{DF}
dp(I|\theta)=\rho(I|\theta)dI,
\end{equation}
where $I$ are the system macroscopic observables that behave as
stochastic variables in an equilibrium situation driven by a set
$\theta$ of control parameters. Let us denote by
$\mathcal{M}_{\theta}$ the compact space constituted by all
admissible values of the macroscopic observables $I$ that are
accessible for a given value $\theta$ of control parameters. Let us
also admit that the probability density $\rho(I|\theta)$ is
everywhere finite and differentiable, and obeys the following
boundary conditions for every
$I_{b}\in\partial\mathcal{M}_{\theta}$:
\begin{equation}\label{boundary}
\lim_{I\rightarrow I_{b}}\rho(I|\theta)=\lim_{I\rightarrow
I_{b}}\frac{\partial}{\partial I^{i}}\rho(I|\theta)=0.
\end{equation}
Let us introduce the \textit{differential generalized forces}
$\eta_{i}(I|\theta)$ as follows:
\begin{equation}\label{DGF}
\eta_{i}(I|\theta)=-\frac{\partial}{\partial
I^{i}}\log\rho(I|\theta).
\end{equation}
By definition, the differential generalized forces
$\eta_{i}(I|\theta)$ vanish in those stationary points $\bar{I}$
where the probability density $\rho(I|\theta)$ exhibits its local
maxima or its local minima. The global (local) maximum of the
probability density is commonly regarded as a \textit{stable
(metastable) equilibrium configuration} in the framework of large
thermodynamic systems. In general, the differential generalized
forces $\eta_{i}(I|\theta)$ characterize the deviation of a given
point $I\in\mathcal{M}_{\theta}$ from these local equilibrium
configurations. As stochastic variables, the expectation values of
the differential generalized forces $\eta_{i}=\eta_{i}(I|\theta)$
identically vanish:
\begin{equation}\label{equi.cond}
\left\langle \eta_{i}\right\rangle =0,
\end{equation}
and these quantities also obey the \textit{fundamental and the
associated fluctuation theorems}:
\begin{equation}
\left\langle \eta_{i}I^{j}\right\rangle
=\delta_{i}^{j},~\left\langle
-\partial_{i}\eta_{j}+\eta_{i}\eta_{j}\right\rangle =0,
\label{fund.assoc}
\end{equation}
where $\partial_{i}A=\partial A/\partial I^{i}$. As already shown in
Ref.\cite{Vel.URSM}, the previous fluctuation relations are directly
derived from the following mathematical identity:
\begin{equation}\label{identity}
\left\langle\partial_{i}
A(I|\theta)\right\rangle=\left\langle\eta_{i}(I|\theta)A(I|\theta)\right\rangle
\end{equation}
substituting the cases $A(I|\theta)=1$, $I^{i}$ and $\eta_{i}$,
respectively. Here, $A(I)$ is a differentiable function of the
macroscopic observables $I$ with definite expectation values
$\left\langle\partial A(I|\theta)/\partial I^{i}\right\rangle$ that
obeys the following the boundary condition:
\begin{equation}
\lim_{I\rightarrow I_{b}}A(I)\rho(I|\theta)=0.
\end{equation}

Fluctuation theorems (\ref{equi.cond}) and (\ref{fund.assoc}) can be
regarded as the counterparts of some known results of
\textit{inference theory} \cite{Fisher,Rao}. To clarify this idea,
let us admit that the probability density $\rho(I|\theta)$ is
everywhere differentiable and finite on the compact space
$\mathcal{P}$ constituted by all admissible values of control
parameters $\theta$. Introducing the \textit{score vectors}
$\upsilon_{\alpha}(I|\theta)$:
\begin{equation}
\upsilon_{\alpha}(I|\theta)=-\frac{\partial}{\partial
\theta^{\alpha}}\log\rho(I|\theta),
\end{equation}
it is possible to obtain the following identities:
\begin{equation}
\left\langle \upsilon_{\alpha}\right\rangle =0,~\left\langle
\upsilon_{\alpha}\hat{\theta}^{\beta}\right\rangle
=-\delta_{\alpha}^{\beta},~\left\langle
-\partial_{\alpha}\upsilon_{\beta}+\upsilon_{\alpha}\upsilon_{\beta}\right\rangle
=0, \label{eq.gen}
\end{equation}
where $\hat{\theta}^{\alpha}=\hat{\theta}^{\alpha}(I)$ denotes any
\textit{unbiased estimator} of the $\alpha$-th control parameter
$\theta^{\alpha}$:
\begin{equation}
\left\langle\hat{\theta}^{\alpha}\right\rangle=\int_{\mathcal{M}_{\theta}}\hat{\theta}^{\alpha}(I)
\rho(I|\theta)dI=\theta^{\alpha},
\end{equation}
and $\partial_{\alpha}A=\partial A/\partial \theta^{\alpha}$. The
previous relations are obtained from the identity:
\begin{equation}
\partial_{\alpha }\left\langle A\left(
I|\theta \right) \right\rangle =\left\langle \partial_{\alpha
}A\left( I|\theta \right) \right\rangle -\left\langle
A(I|\theta)\upsilon _{\alpha }\left( I|\theta \right) \right\rangle,
\end{equation}
where $A(I|\theta)$ is any differentiable function of control
parameters $\theta$ with a definite statistical expectation value:
\begin{equation}
\left\langle
A(I|\theta)\right\rangle=\int_{\mathcal{M}_{\theta}}A(I|\theta)\rho(I|\theta)dI.
\end{equation}
Introducing the inverse matrix $g^{\alpha\beta}(\theta)$ of the
self-correlation matrix $g_{ij}(\theta)$:
\begin{eqnarray}
g_{ij}(\theta)=\left\langle\upsilon_{\alpha}(I|\theta)\upsilon_{\beta}(I|\theta)
\right\rangle \\ \nonumber
=\int_{\mathcal{M}_{\theta}}\left[-\partial_{\alpha}\log\rho(I|\theta)\right]
\left[-\partial_{\beta}\log\rho(I|\theta)\right]\rho(I|\theta)dI
\end{eqnarray}
and the auxiliary quantity
$X^{\alpha}=\delta\hat{\theta}^{\alpha}-g^{\alpha\beta}\upsilon_{\beta}$,
one can compose the positive definite form:
\begin{equation}
\left\langle\left(\lambda_{\alpha}X^{\alpha}\right)^{2}\right\rangle=\left\langle
X^{\alpha}X^{\beta}\right\rangle\lambda_{\alpha}\lambda_{\beta}\geq
0,
\end{equation}
which leads to the positive definition of the matrix:
\begin{equation}\label{CramerRao}
\left\langle\delta
\hat{\theta}^{\alpha}\delta\hat{\theta}^{\beta}\right\rangle-g^{\alpha\beta}(\theta)\succeq
0.
\end{equation}
Eq.(\ref{CramerRao}) is the known \textit{Cramer-Rao theorem} of
inference theory \cite{Fisher,Rao} that imposes an inferior bound to
the efficiency of unbiased estimators $\hat{\theta}^{\alpha}$, where
the matrix (\ref{CramerRao}) is referred to as the Fisher's
information matrix. Considering the inverse $M^{ij}(\theta)$ of the
self-correlation matrix of the differential generalized forces:
\begin{equation}
M_{ij}(\theta)=\left\langle\eta_{i}(I|\theta)\eta_{j}(I|\theta)\right\rangle,
\end{equation}
it is possible to obtain the following matrical inequalities:
\begin{equation}
\left\langle\delta I^{i}\delta
I^{j}\right\rangle-M^{ij}(\theta)\succeq 0.
\end{equation}
Clearly, this last result is a counterpart of the Cramer-Rao theorem
(\ref{CramerRao}) in the framework of fluctuation theory.
Introducing the gradient operators
$\partial_{i}\rightarrow\mathbf{\nabla}_{I}$ and
$\partial_{\alpha}\rightarrow\mathbf{\nabla}_{\theta}$, the diadic
products $\mathbf{A}\cdot \mathbf{B}=A_{i}B_{j}\mathbf{e}^{i}\cdot
\mathbf{e}^{j}$ and $\mathbf{\xi}\cdot
\mathbf{\psi}=\xi_{\alpha}\psi_{\beta}\mathbf{\epsilon}^{\alpha}\cdot
\mathbf{\epsilon}^{\beta}$ and the delta Kroneker
$\delta^{i}_{j}\rightarrow\mathbf{1}_{I}$ and
$\delta^{\alpha}_{\beta}\rightarrow\mathbf{1}_{\theta}$, the
underlying analogy between fluctuation theory and inference theory
is summarized in Table \ref{Duality}.

\begin{table}[tbp] \centering%
\begin{tabular}{|c|c|}
\hline\hline \textbf{Inference theory} & \textbf{Fluctuation theory}
\\ \hline\hline
\multicolumn{1}{|c|}{$\mathbf{\upsilon }\left( \mathcal{I}|\theta \right) =-%
\mathbf{\nabla }_{\theta }\log \rho \left( \mathcal{I}|\theta \right) $} & $%
\mathbf{\eta }\left( I|\theta \right) =-\mathbf{\nabla }_{I}\log
\rho \left(
I|\theta \right) $ \\
$\left\langle \mathbf{\upsilon }\left( \mathcal{I}|\theta \right)
\right\rangle =0$ & $\left\langle \mathbf{\eta }\left( I|\theta
\right)
\right\rangle =0$ \\
$\left\langle \mathbf{\upsilon }\left( \mathcal{I}|\theta \right)
\cdot \delta \mathbf{\hat{\theta}}\right\rangle
=-\mathbf{1}_{\theta}$ & $\left\langle
\mathbf{\eta }\left( I|\theta \right) \cdot \delta \mathbf{I}\right\rangle =%
\mathbf{1}_{I}$ \\
$\left\langle \mathbf{\nabla }_{\theta }\cdot \mathbf{\upsilon
}\left(
\mathcal{I}|\theta \right) \right\rangle =\left\langle \mathbf{\upsilon }%
\left( \mathcal{I}|\theta \right) \cdot \mathbf{\upsilon }\left( \mathcal{I}%
|\theta \right) \right\rangle $ & $\left\langle \mathbf{\nabla
}_{I}\cdot
\mathbf{\eta }\left( I|\theta \right) \right\rangle =\left\langle \mathbf{%
\eta }\left( I|\theta \right) \cdot \mathbf{\eta }\left( I|\theta
\right) \right\rangle $ \\ \hline
\end{tabular}%
\caption{Analogy between inference theory and fluctuation
theory.}\label{Duality}%
\end{table}%

\subsection{Fluctuation theorems for systems in contact with an environment}

Let us apply the fluctuation relations (\ref{equi.cond}) and
(\ref{fund.assoc}) to equilibrium distribution functions of
classical statistical mechanics. As discussed elsewhere, classical
fluctuation theory starts from the Einstein postulate
\cite{landau,Reichl}:
\begin{equation}\label{EP}
dp(I|\theta)=A\exp\left[S(I|\theta)\right]dI,
\end{equation}
which allows to relate the differential generalized forces with the
entropy $S(I|\theta)$:
\begin{equation}
\mathbf{\eta}(I|\theta)=-\mathbf{\nabla}_{I}S(I|\theta).
\end{equation}
Let us assume that the entropy $S(I|\theta)$ can be decomposed into
two additive terms:
\begin{equation}\label{SDecomp}
S(I|\theta)=S(I)+S^{\omega}(I|\theta),
\end{equation}
where $S(I)$ is the entropy of an isolated system, while
$S^{\omega}(I|\theta)$ is the contribution of the total entropy
$S(I|\theta)$ when this system is put in thermodynamic equilibrium
with a certain environment. Such a decomposition leads to the
following distribution function:
\begin{equation}\label{GenEns}
dp(I|\theta)=\omega(I|\theta)\Omega(I)dI,
\end{equation}
where the probabilistic weight
$\omega(I|\theta)\sim\exp\left[S(I|\theta)\right]$ arises here as a
formal extension of the Boltzmann-Gibbs distribution (\ref{BGD}).
The differential generalized forces can be rephrased as follows:
\begin{equation}
\eta_{i}(I|\theta)=\beta^{\omega}_{i}(I|\theta)-\beta_{i}(I|\theta),
\end{equation}
where $\beta^{\omega}_{i}(I|\theta)$ denotes the environmental
generalized forces:
\begin{equation}\label{env.ctrl.variab}
\beta^{\omega}_{i}(I|\theta)=-\frac{\partial
S^{\omega}(I|\theta)}{\partial I^{i}},
\end{equation}
and $\beta_{i}(I)$ is the system generalized forces:
\begin{equation}\label{mic.conj}
\beta_{i}(I)=\frac{\partial S (I)}{\partial I^{i}}.
\end{equation}
Eq.(\ref{equi.cond}) drops to the \textit{equilibrium thermodynamic
conditions} in the form of statistical expectation values:
\begin{equation}\label{eq}
\left\langle\beta_{i}(I)\right\rangle=\left\langle\beta^{\omega}_{i}(I)\right\rangle.
\end{equation}
The relevance of the fundamental fluctuation theorem in
Eq.(\ref{eq.gen}) with the existence of \textit{complementary
relations of statistical mechanics} has been previously discussed in
an extensive way \cite{Vel.URSM}. The complementary fluctuation
theorem is the generalization of the identity (30) obtained in
Ref.\cite{Vel.Geo}, which identifies the expectation value of the
response matrix of the differential generalized forces $\zeta_{ij}$:
\begin{equation}\label{resp.forces}
\zeta_{ij}=\frac{\partial \eta_{j}}{\partial I^{i}}
\end{equation}
and its self-correlation matrix $\left\langle \delta\eta
_{i}\delta\eta_{j}\right\rangle$. This general fluctuation theorem
is a particular expression of the \textit{Le Chatellier-Braun
principle} \cite{Reichl}: the response of a stable system to the
action from outside must be a weakened resistance to this action.
This behavior is manifested as the positive definite character of
the response matrix $\zeta_{ij}$ in the differential generalized
forces $\eta_{i}$, which can be inferred from the positive definite
character of the self-correlation matrix $\left\langle \delta\eta
_{i}\delta\eta_{j}\right\rangle$. Fluctuation theorems
(\ref{fund.assoc}) constitute the most general extension of some
known results of classical fluctuation theory. For example,
component of the fundamental fluctuation theorem involving the
differential inverse temperature and the volume $V$ of a fluid
system:
\begin{equation}
\left\langle\delta V\left(\frac{1}{T^{\omega}}-\frac{1}{T}\right)%
\right\rangle=0,
\end{equation}
as well as the term of the complementary fluctuation theorem
involving the system internal energy $U$ and its temperature $T$:
\begin{equation}
\left\langle\frac{\partial}{\partial
U}\left(\frac{1}{T^{\omega}}-\frac{1}{T}\right)%
\right\rangle=\left\langle\left(\frac{1}{T^{\omega}}%
-\frac{1}{T}\right)^{2}\right\rangle
\end{equation}
drop to the familiar expressions \cite{landau}:
\begin{equation}
\left\langle\delta V\delta T \right\rangle=0,\quad
T^{2}/C_{V}=\left\langle \delta T^{2}\right\rangle
\end{equation}
after considering the first-order approximation discussed below and
the constant character of the environmental temperature $T^{\omega}$
(canonical ensemble), where $C_{V}$ is the system heat capacity at
constant volume.

Let us employ the exact theorems (\ref{fund.assoc}) to arrive at a
suitable extension of the energy-temperature fluctuation relation
(\ref{unc}). Hereafter, we shall admit a first-order approximation
where the expectation value and the fluctuations of a differentiable
function $A\left( I\right) $ can be expressed as follows:
\begin{equation}
\left\langle A(I)\right\rangle\simeq A(\bar{I}),\quad\delta A\left(
I\right) \simeq\frac{\partial A\left( \bar{I}\right)}{\partial
I^{i}} \delta I^{i}, \label{first.ord}
\end{equation}
where $\bar{I}$ represents the most likely state. Additionally, we
shall omit the dependence of the thermodynamic functions on the
macroscopic observables $I$ to adopt a more simple notation in the
mathematical expressions. Using the approximation rules
(\ref{first.ord}), the fluctuation of the generalized forces
$\beta_{i}$ can be expressed in terms of the entropy Hessian
$H_{ij}$:
\begin{equation}\label{hessian}
H_{ij} =\frac{\partial\beta_{j}}{\partial
I^{i}}=\frac{\partial^{2}S}{\partial I^{i}\partial I^{j}}
\end{equation}
as follows:
\begin{equation}
\delta\beta_{i}^{s}=H_{ik}\delta I^{k}. \label{hh}
\end{equation}
Using these approximations, one can rephrased the fundamental and
the associated fluctuation theorems (\ref{fund.assoc}) as follows:
\begin{eqnarray}\label{fund.theorem}
  \chi^{ij}=\left\langle\delta I^{i}\delta
I^{j}\right\rangle+\chi^{ik}\left\langle\delta
\beta^{\omega}_{k}\delta I^{j}\right\rangle,\\
 \zeta_{ij} =\left\langle
\delta\eta_{i}\delta\eta_{j}\right\rangle,\label{assoc.theorem}
\end{eqnarray}
where the response matrix $\chi^{ij}=-H^{ij}$ is given by the
inverse $H^{ij}$ of the entropy Hessian (\ref{hessian}). The
previous reasonings support the existence of a third equilibrium
fluctuation theorem that has no counterpart in the framework of the
Boltzmann-Gibbs distributions (\ref{BGD}). This is the
\textit{associated fluctuation theorem}:
\begin{equation}\label{complementary}
\left\langle
\delta\beta_{i}^{\omega}\delta\beta_{k}^{\omega}\right\rangle
\left\langle \delta I^{k}\delta I^{j}\right\rangle =\left\langle
\delta \beta_{i}^{\omega}\delta I^{k}\right\rangle \left\langle
\delta\beta _{k}^{\omega}\delta I^{j}\right\rangle,
\end{equation}
which trivially vanishes for the statistical ensemble (\ref{BGD})
since it involves the self-correlation matrix of the environmental
control variables $\left\langle \delta\beta_{i}^{\omega
}\delta\beta_{j}^{\omega}\right\rangle $. General speaking, the
obtaining of the correlation matrix $\left\langle
\delta\beta_{i}^{\omega}\delta I^{j}\right\rangle $ demands to
perform \textit{simultaneous measurements} of the macroscopic
observables $\left\{I^{i}\right\}$ and the environmental control
variables $\left\{\beta^{\omega}_{i}\right\}$, which are very
difficult to carry out in the practice. Such a difficulty can be
overcome using the associated fluctuation theorem
(\ref{complementary}), which allows an indirect determination of the
correlation matrix $\left\langle \delta\beta_{i}^{\omega }\delta
I^{j}\right\rangle $ performing independent measurements of the
self-correlation matrixes $\left\langle
\delta\beta_{i}^{\omega}\delta\beta_{j}^{\omega}\right\rangle $ and
$\left\langle \delta I^{i}\delta I^{j}\right\rangle$. The proof of
this theorem starts from introducing the Hessian $F_{ij}$:
\begin{equation}
F_{ij}=\frac{\partial \beta^{\omega}_{i}}{\partial
I^{j}}=-\frac{\partial^{2}S^{\omega}(I|\theta)}{\partial
I^{i}\partial I^{j}}.
\end{equation}
Using the first-order approximation:
\begin{equation}
\delta\beta_{i}^{\omega}=F_{ij}\delta I^{j},
\end{equation}
one can show the following relations:
\begin{equation}
\left\langle
\delta\beta_{i}^{\omega}\delta\beta_{j}^{\omega}\right\rangle
=F_{jk} \left\langle \delta\beta_{i}^{\omega}\delta
I^{k}\right\rangle, \left\langle \delta\beta_{i}^{\omega}\delta
I^{j}\right\rangle =F_{ik}\left\langle \delta I^{k}\delta
I^{j}\right\rangle,
\end{equation}
which can be easily combined to obtain the desirable result
(\ref{complementary}).

Let us rewrite the equilibrium fluctuation theorems
(\ref{fund.theorem})-(\ref{complementary}) in terms of the ordinary
variables considered in thermodynamics, $\left( U,\beta\right) $ and
$\left( X,Y\right) $. The counterpart of the fluctuation theorem
(\ref{fund.theorem}) within the Boltzmann-Gibbs distribution
(\ref{BGD}) can be rewritten in compact matrix form as follows:
\begin{equation}
\mathcal{R}=\mathcal{C},  \label{can1}
\end{equation}
where the response and the self-correlation matrices $\mathcal{R}$
and $\mathcal{C}$ are given by:
\begin{equation}
\mathcal{R}=-\left(
\begin{array}{cc}
\partial_{\beta}\left\langle \mathcal{H}\right\rangle & \partial_{\beta}\left(
\beta\left\langle X\right\rangle \right) \\
\partial_{Y}\left\langle \mathcal{H}\right\rangle & \beta\partial_{Y}\left\langle
X\right\rangle%
\end{array}
\right), \mathcal{C}=\left(
\begin{array}{cc}
\left\langle \delta Q^{2}\right\rangle & \beta\left\langle \delta Q\delta
X\right\rangle \\
\beta\left\langle \delta X^{T}\delta Q\right\rangle &
\beta^{2}\left\langle
\delta X^{T}\delta X\right\rangle%
\end{array}
\right).%
\label{asso.FR}
\end{equation}
Here, we have adopted the following matrix conventions:
\begin{eqnarray}
  X= \left(%
\begin{array}{cccc}
 X^{1} & X^{2} & \ldots & X^{n}  \\
\end{array}
\right),
\partial_{X}= \left(%
\begin{array}{cccc}
 \partial_{X^{1}} & \partial_{X^{2}} & \ldots & \partial_{X^{n}}  \\
\end{array}
\right), \\
  Y^{T}= \left(%
\begin{array}{cccc}
 Y_{1} & Y_{2} & \ldots & Y_{n}  \\
\end{array}
\right),
\partial_{Y}^{T}= \left(%
\begin{array}{cccc}
 \partial_{Y_{1}} & \partial_{Y_{2}} & \ldots & \partial_{Y_{n}}  \\
\end{array}
\right),
\end{eqnarray}
where $A^{T}$ denotes the transpose operation. Note that the
generalized forces $Y$ and their differential operators
$\partial_{Y}$ represent column vectors while $\left\langle \delta
X^{T}\delta X\right\rangle$ and $\partial_{Y}\left\langle
X\right\rangle$ are $n\times n$ square matrices. Besides,
$\mathcal{H}=U+XY$ is the Enthalpy (we employ here this notation to
avoid any ambiguity with the magnetic field $H$), and $\delta
Q=\delta U+\delta X Y$ the amount of heat absorbed or transferred by
the system from its environment at the equilibrium, where
$\left\langle\delta Q\right\rangle\equiv0$. It is important to bear
in mind that the Enthalpy fluctuation $\delta \mathcal{H}=\delta Q$
within the Boltzmann-Gibbs distribution (\ref{BGD}). Such a
relationship does not hold when the fluctuations of the generalized
forces $Y$ are taken into consideration, where $\delta
\mathcal{H}=\delta Q+X\delta Y$. The matrix form of the fluctuation
theorem (\ref{can1}) in the symmetric representation using the
conjugated thermodynamic variables $\left( U,\beta\right) $ and
$\left( X,\xi\right) $ with $\xi=\beta Y$ reads as follows:
\begin{equation}
\chi=C,  \label{can2}
\end{equation}
where the response and the self-correlation matrices $\chi$ and $C$
are given by:
\begin{equation}
\chi=-\left(
\begin{array}{cc}
\partial_{\beta}\left\langle U\right\rangle & \partial_{\beta}\left\langle
X\right\rangle \\
\partial_{\xi}\left\langle U\right\rangle & \partial_{\xi}\left\langle
X\right\rangle%
\end{array}
\right) , C=\left(
\begin{array}{cc}
\left\langle \delta U^{2}\right\rangle & \left\langle \delta U\delta
X\right\rangle \\
\left\langle \delta X^{T}\delta U\right\rangle & \left\langle \delta
X^{T}\delta
X\right\rangle%
\end{array}
\right).
\end{equation}
Representations (\ref{can1}) and (\ref{can2}) are related by the
following transformation rules:
\begin{equation}\label{T1}
\mathcal{C}=\mathcal{T}C\mathcal{T}^{T},~\mathcal{R}=\mathcal{T}\chi\mathcal{%
T}^{T},
\end{equation}
where the transformation matrix $\mathcal{T}$ is given by:%
\begin{equation}
\mathcal{T}=\left(
\begin{array}{cc}
\hat{1} & Y \\
0 & \beta%
\end{array}
\right) ,
\end{equation}
with $\hat{1}$ being the $n\times n$ unitary matrix. Thus, the
fundamental fluctuation theorem (\ref{fund.theorem}) is rewritten as
follows:
\begin{equation}\label{FFT.Th}
\chi=C+\chi D\rightarrow\mathcal{R}=\mathcal{C}+\mathcal{RD},
\label{gen.fdr}
\end{equation}
where the correlation matrices $D$ and $\mathcal{D}$ characterizing
the environmental feedback effects
\begin{eqnarray}
  D=\left(
\begin{array}{cc}
\left\langle \delta\beta^{\omega}\delta U\right\rangle &
\left\langle
\delta\beta^{\omega}\delta X\right\rangle \\
\left\langle \delta\xi^{\omega}\delta U\right\rangle & \left\langle
\delta
\xi^{\omega}\delta X\right\rangle%
\end{array}
\right)\mbox{ and } \\
  \mathcal{D}=\left(
\begin{array}{cc}
\left\langle \delta\beta^{\omega}\delta Q\right\rangle &
\beta\left\langle
\delta\beta^{\omega}\delta X\right\rangle \\
\left\langle \delta Y^{\omega}\delta Q\right\rangle &
\beta\left\langle
\delta Y^{\omega}\delta X\right\rangle%
\end{array}
\right) \label{A2.FR}
\end{eqnarray}
are related by the transformation rule:
\begin{equation}\label{T2}
\mathcal{D}=\left( \mathcal{T}^{T}\right) ^{-1}D\mathcal{T}^{T}.
\end{equation}
Here, the quantities $\left( \beta^{\omega},Y^{\omega}\right) $
denote the environmental control variables in the usual
representation of thermodynamics.

The complementary fluctuation theorem (\ref{assoc.theorem}) is
written in compact matrix form as follows:
\begin{equation}
N=F,
\end{equation}
where $N$ and $F$ are the response and self-correlation matrices of
differential generalized forces $\eta =\beta ^{\omega }-\beta $ and
$\eta _{X}=\xi ^{\omega }-\xi $ in the symmetric representation:
\begin{equation}
N=\left(
\begin{array}{cc}
\partial _{U}\eta  & \partial _{U}\eta^{T} _{X} \\
\partial^{T} _{X}\eta  & \partial^{T} _{X}\eta^{T} _{X}%
\end{array}%
\right) ,F=\left(
\begin{array}{cc}
\left\langle \delta \eta ^{2}\right\rangle  & \left\langle \delta
\eta
\delta \eta^{T} _{X}\right\rangle  \\
\left\langle \delta \eta _{X}\delta \eta \right\rangle  &
\left\langle \delta \eta _{X}\delta \eta^{T} _{X}\right\rangle
\end{array}%
\right) .
\end{equation}%
Using the transformation rules:
\begin{equation}\label{T3}
\mathcal{N}=\left( \mathcal{T}^{T}\right) ^{-1}N\mathcal{T}^{-1},\mathcal{F}%
=\left( \mathcal{T}^{T}\right) ^{-1}F\mathcal{T}^{-1},
\end{equation}%
this fluctuation theorem can be rewritten as follows:
\begin{equation}\label{AFT.Th}
N=F\rightarrow \mathcal{N}=\mathcal{F},
\end{equation}%
where $\mathcal{N}$ and $\mathcal{F}$ are their respective
expressions in the representation of thermodynamics:
\begin{eqnarray}
\mathcal{N}=\left(
\begin{array}{cc}
\partial _{U}\eta  & \partial _{U}\mathcal{Y}^{T} \\
T\left( \partial^{T} _{X}\eta -Y\partial _{U}\eta \right)  & T\left(
\partial^{T}_{X}\mathcal{Y}^{T}-Y\partial _{U}\mathcal{Y}^{T}\right)
\end{array}%
\right) ,\\
\mathcal{F}=\left(
\begin{array}{cc}
\left\langle \delta \eta ^{2}\right\rangle  & \left\langle \delta
\eta
\delta \mathcal{Y}\right\rangle  \\
\left\langle \delta \mathcal{Y}\delta \eta \right\rangle  &
\left\langle \delta \mathcal{Y}^{2}\right\rangle
\end{array}%
\right),
\end{eqnarray}%
with $\mathcal{Y}=Y^{\omega }-Y$  being the differential generalized
force conjugated with the observable $X$. Using the same procedure,
the associated fluctuation theorem (\ref{complementary}) can be
written as follows:
\begin{equation}
BC=D^{2},
\end{equation}
where $B$ is the self-correlation matrix of the environmental
control variables in the symmetric representation:
\begin{equation}\label{BB}
B=\left(
\begin{array}{cc}
\left\langle (\delta\beta^{\omega})^{2}\right\rangle & \left\langle
\delta\beta^{\omega}(\delta \xi^{\omega})^{T}\right\rangle \\
\left\langle \delta\xi^{\omega}\delta\beta^{\omega}\right\rangle &
\left\langle \delta \xi^{\omega}(\delta
\xi^{\omega})^{T}\right\rangle%
\end{array}
\right).
\end{equation}
Considering the transformation rule:
\begin{equation}\label{T4}
\mathcal{B}=\left( \mathcal{T}^{T}\right) ^{-1}B\mathcal{T}^{-1},
\end{equation}%
this fluctuation theorem can be rewritten as follows:
\begin{equation}\label{CFT.Th}
\mathcal{BC}=\mathcal{D}^{2},
\end{equation}
where the self-correlation matrix of the environmental control
variables $\mathcal{B}$ in the new representation is given by:
\begin{equation}\label{BBT}
\mathcal{B}=\left(
\begin{array}{cc}
\left\langle (\delta\beta^{\omega})^{2}\right\rangle & \left\langle
\delta\beta^{\omega}(\delta Y^{\omega})^{T}\right\rangle \\
\left\langle \delta Y^{\omega}\delta\beta^{\omega}\right\rangle &
\left\langle \delta Y^{\omega}(\delta
Y^{\omega})^{T}\right\rangle%
\end{array}
\right).
\end{equation}

The mathematical expressions of equilibrium fluctuation theorems
(\ref{fund.theorem})-(\ref{complementary}) manifest the non
preference of the thermodynamic description on a given macroscopic
observable. This feature differs from the character of their
respective expressions (\ref{FFT.Th}), (\ref{AFT.Th}) and
(\ref{CFT.Th}), which explicitly attributes a preference on the
energy and its conjugated quantity, the temperature. This second
representation allows the matching of the current approach to the
thermodynamic quantities obtained from the experiment. However, it
is always easier to perform calculations using the symmetric
representation, and after, to use the respective transformation
rules to refer the response and correlation matrices in the second
representation.

Fluctuation theorem (\ref{fund.assoc}) has been employed to obtain
the extension of conventional fluctuation theorems relating the
correlation functions of macroscopic fluctuations with response
functions. However, one can also rewrite this last fluctuation
theorem to emphasize the complementary character of conjugated
thermodynamic quantities, which is precisely its main physical
content. Using the transformation rule:
\begin{equation}\label{T5}
\mathcal{S}=\left( \mathcal{T}^{T}\right) ^{-1}S\mathcal{T}^{T},
\end{equation}
one can rewrite the original form of the fundamental fluctuation
theorem (\ref{fund.assoc}) as follows:
\begin{equation}
S=\hat{1}\rightarrow \mathcal{S}=\hat{1},
\end{equation}
where $\hat{1}$ is the unitary matrix, while $S$ and $\mathcal{S}$
represent the correlation matrix between differential generalized
forces and the macroscopic observables in these two representations:
\begin{equation}
S=\left(
\begin{array}{cc}
\left\langle \delta\eta\delta U\right\rangle & \left\langle
\delta\beta\delta X\right\rangle \\
\left\langle \delta\eta_{X}\delta U\right\rangle & \left\langle
\delta
\eta_{X}\delta X\right\rangle%
\end{array}
\right), \mathcal{S}=\left(
\begin{array}{cc}
\left\langle \delta\eta\delta Q\right\rangle & \beta\left\langle
\delta\eta\delta X\right\rangle \\
\left\langle \delta \mathcal{Y}\delta Q\right\rangle &
\beta\left\langle
\delta \mathcal{Y}\delta X\right\rangle%
\end{array}
\right).
\end{equation}

\section{Application of the present approach}\label{application}

\subsection{Relationship among anomalous response and phase transitions}

Anomalous response (or states where the response matrix
$\mathcal{R}$ is non positive definite) are intimately related to
the occurrence of phase transitions. Indeed, a phase transition is
the manifestation of a \textit{thermodynamic instability}, and
precisely, macrostates with anomalous response are always unstable
within the Boltzmann-Gibbs statistics (\ref{BGD}). The best known
example is the relation between negative heat capacities and the
occurrence of a temperature driven discontinuous phase transition
\cite{Lyn3,gro1}. The same kind of relationship also appears in
other anomalous response functions,
\begin{figure}[t]
\begin{center}
\includegraphics[
height=4.0in, width=5.0in ]{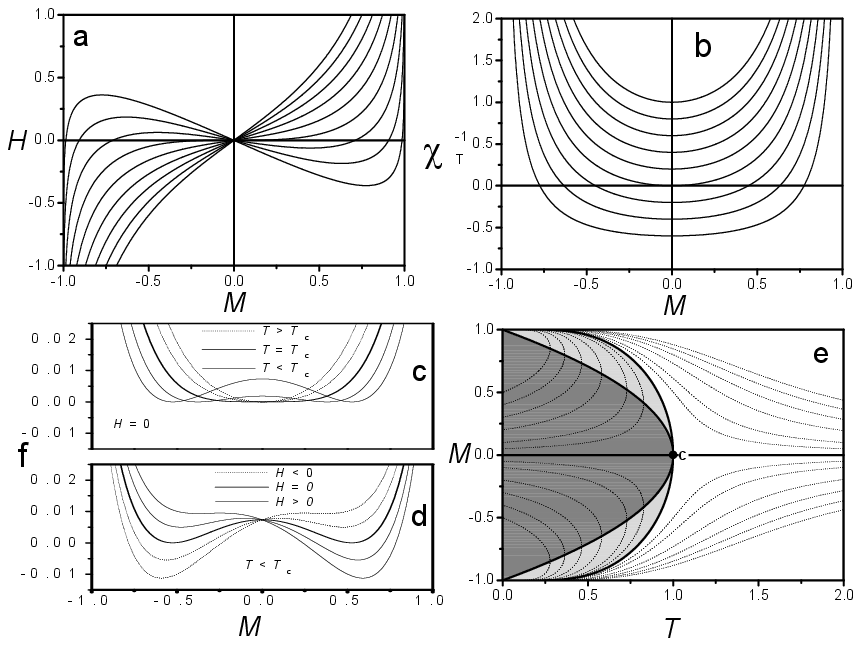}
\end{center}
\caption{Incidence of anomalous diamagnetic states on the
thermodynamic description of the ferromagnetic Weiss model. Panel
\textbf{a}: Isotherms in $H-M$ plane. Panel \textbf{b}: inverse of
the isothermal magnetic susceptibility curve $\chi_{T}$. Panels
\textbf{c} and \textbf{d}: Behavior of minima and maxima of the
function $f(M;T,H)$. Panel \textbf{e}: Phase diagram and
magnetization curves at constant magnetic field (dotted lines).}
\label{Weiss.eps}
\end{figure}
as it is evidenced in the thermodynamic description of the
ferromagnetic Weiss model \cite{Stanley} shown in
FIG.\ref{Weiss.eps}. The magnetization per particle $m=m(H)$
dependence on the external magnetic field $H$ can be rewritten as
follows:
\begin{equation}
m(H)=\mu\tanh\left[\beta\mu(H+\alpha m)\right]\rightarrow
H=\frac{T}{\mu}\tanh^{-1}\left(\frac{m}{\mu}\right)-\alpha m
\end{equation}
to reveal the existence of states with negative isothermal
susceptibilities $\chi_{T}<0$:
\begin{equation}
N\chi^{-1}_{T}=\frac{T}{\mu^{2}-m^{2}}-\alpha
\end{equation}
for temperatures below the critical temperature
$T_{c}=\alpha\mu^{2}$ of the ferromagnetic transition (panels
\textbf{a} and \textbf{b}). Here, $\mu$ is the magnetic moment and
$\alpha$ the molecular field parameter. The unstable character of
these diamagnetic states within the Boltzmann-Gibbs distributions
can be inferred from the fluctuation relation
$\chi_{T}=\beta\left\langle\delta M^{2}\right\rangle$, which is only
compatible with nonnegative susceptibilities. Both the appearance of
a non-vanishing spontaneous magnetization (panel \textbf{c}) and the
sudden jump of magnetization with a small varying of the external
magnetic field $H$ at the value $H=0$ (panel \textbf{d}) are direct
consequences of the existence of these anomalous states. This
relation can be observed in the calculation of the Helmholtz
potential $H(T,H)=\min_{M} f(M;T,H)$, where $f(M;T,H)$ is given by:
\begin{equation}
f(M;T,H)=-\frac{1}{2N}\alpha
M^{2}+T\int\tanh^{-1}\left(\frac{M}{N\mu}\right)\frac{dM}{\mu}-HM.
\end{equation}
Here, the two minima of $f(M;T,H)$ for $T<T_{c}$ correspond to the
stable and metastable states with $\chi_{T}>0$, and its local
maximum, an unstable state with $\chi_{T}<0$. The distinction among
stable (white region), metastable (light gray region) and unstable
(dark gray region) states leads to the phase diagram shown in panel
\textbf{e}. Since the critical point (\textbf{C}) is a state of
\textit{marginal stability} located at the boundary of the unstable
region, the occurrence of a continuous phase transition is also
associated with a region of anomalous response.

\subsection{A special condition of thermodynamic stability}

Diamagnetic states observed below the critical temperature of the
ferromagnetic transition are unstable for the particular equilibrium
situation considered in the previous example: a magnetic system
under the influence of constant environmental temperature and
constant external magnetic field. The same states, however, could be
stable in other equilibrium situations. To illustrate this last
possibility, one should obtain the particular expression of the
fundamental fluctuation theorem (\ref{fund.theorem}) for a magnetic
system with internal energy $U$, total magnetization $M$ and
Enthalpy $\mathcal{H}=U-HM$:
\begin{eqnarray}
\mathcal{R}=\left(
\begin{array}{cc}
T^{2}C_{H} & T\left( \partial M/\partial T\right) _{H}-M \\
\left( \partial \mathcal{H}/\partial H\right) _{T} & \beta\chi_{T}%
\end{array}
\right) ,\label{magnetic}\\
\nonumber \mathcal{C}=\left(
\begin{array}{cc}
\left\langle \delta Q^{2}\right\rangle & \beta\left\langle \delta
Q\delta
M\right\rangle \\
\beta\left\langle \delta M\delta Q\right\rangle &
\beta^{2}\left\langle
\delta M^{2}\right\rangle%
\end{array}
\right) ,  \mathcal{D}=\left(
\begin{array}{cc}
\left\langle \delta\beta^{\omega}\delta Q\right\rangle &
\beta\left\langle
\delta\beta^{\omega}\delta M\right\rangle \\
-\left\langle \delta H^{\omega}\delta Q\right\rangle &
-\beta\left\langle
\delta H^{\omega}\delta M\right\rangle%
\end{array}
\right).
\end{eqnarray}
Here, $C_{H}=\left( \partial \mathcal{H}/\partial T\right) _{H}$ is
the heat capacity at constant magnetic field and $\chi_{T}=\left(
\partial M/\partial H\right) _{T}$ the isothermal magnetic
susceptibility, where the symmetry of the response matrix
$\mathcal{R}$ leads to the thermodynamical identity:
\begin{equation}
\left( \frac{\partial \mathcal{H}}{\partial H}\right) _{T}=T\left( \frac{\partial M}{%
\partial T}\right) _{H}-M.
\end{equation}
Admitting that $\chi_{T}$ is the only anomalous response function,
one can restrict the analysis to an equilibrium situation where the
environmental inverse temperature $\beta^{\omega}$ takes a constant
value $\beta$, but the external magnetic field $H^{\omega}$
undergoes a non-vanishing magnetic feedback effect
$\left\langle\delta H^{\omega}\delta M\right\rangle$. This effect
naturally arises when the source of the external magnetic field
$H^{\omega}$ is disturbed by the magnetic influence of the system.
The simplest way to account for this type of situation is when
$H^{\omega}$ undergoes small fluctuations around its mean value $H$
coupled to the total system magnetization:
\begin{equation}  \label{ans.mag}
H^{\omega}=H-\lambda\delta M/N,
\end{equation}
where $N$ is the system size, and $\lambda$, a coupling constant
characterizing the system-environment magnetic interaction. For this
particular equilibrium situation, the fluctuation relation involving
the isothermal magnetic susceptibility $\chi_{T}$ :
\begin{equation}
\beta\chi_{T} =\beta^{2}\left\langle \delta M^{2}\right\rangle
+\left[ T\left( \partial M/\partial T\right) _{H}-M\right]
\beta\left\langle
\delta\beta^{\omega}\delta M\right\rangle  \nonumber \\
-\beta^{2}\chi_{T}\left\langle \delta H^{\omega}\delta M\right\rangle
\label{iso.sus}
\end{equation}
drops to:
\begin{equation}
\chi_{T}=\beta\left\langle \delta M^{2}\right\rangle
-\beta\chi_{T}\left\langle \delta H^{\omega}\delta M\right\rangle.
\end{equation}
Clearly, this expression is very similar to the energy-temperature
fluctuation relation (\ref{unc}), which only involves conjugated
quantities as the system magnetization $M$ and the external magnetic
field $H^{\omega}$. This relation can be rewritten as:
\begin{equation}  \label{iso.sus2}
\beta\left\langle \delta M^{2}\right\rangle
=\frac{\chi_{T}}{1+\lambda\chi _{T}/N}
\end{equation}
after using the ansatz (\ref{ans.mag}). Eq.(\ref{iso.sus2}) can be
employed to obtain the self-correlation function of the external
magnetic field $H^{\omega}$:
\begin{equation}  \label{iso.sus3}
\left\langle (\delta H^{\omega})^{2}\right\rangle=\frac{1}{N^{2}}%
\lambda^{2}\left\langle \delta M^{2}\right\rangle=\frac{1}{N\beta}\lambda^{2}%
\frac{\chi_{T}/N}{1+\lambda\chi _{T}/N}.
\end{equation}
For extensive systems, the isothermal susceptibility $\chi_{T}$
usually grows with $N$ as $\chi_{T}\propto N$. Consequently, the
self-correlation functions of the system magnetization and the
external magnetic field behave as $ \left\langle \delta
M^{2}\right\rangle\propto N$ and $\left\langle (\delta
H^{\omega})^{2}\right\rangle\propto 1/N$. Since the fluctuations of
the external magnetic field $H^{\omega}$ vanish in the thermodynamic
limit $N\rightarrow\infty$, the present equilibrium seems to be very
similar to the conventional situation where the external magnetic
field is constant. However, both the fluctuating behavior described
by the expressions (\ref {iso.sus2}) and (\ref{iso.sus3}), as well
as stability condition (\ref {mag.cond}) depend on the coupling
constant $\lambda$. Indeed, the coupling constant $\lambda$ can be
appropriately chosen to force the thermodynamic stability of
anomalous diamagnetic states $\chi_{T}<0$. Since the
self-correlation function of the system magnetization $\left\langle
\delta M^{2}\right\rangle$ is nonnegative, anomalous diamagnetic
states $\chi_{T}<0$ are thermodynamically stable when the condition:
\begin{equation}  \label{mag.cond}
\lambda+N/\chi_{T}>0
\end{equation}
holds. This last result constitutes the magnetic counterpart of the
Thirring's constraint \cite{thir}:
\begin{equation}
C_{B}<\left|C\right|
\end{equation}
reobtained in Ref.\cite{Vel.Thirring} from the energy-temperature
fluctuation relation (\ref{unc}) to force the thermal stability of
states with negative heat capacities $C<0$, with $C_{B}$ being the
heat capacity of a finite thermostat.

\subsection{Monte Carlo study of 2D Ising model}

The consequences derived from the previous analysis are easily
tested with the help of Monte Carlo (MC) simulations. For example,
let us now consider the 2D Ising model on the square lattice
$L\times L$ with periodic boundary condition:
\begin{equation}
U=-\sum_{\left\langle ij\right\rangle }s_{i}s_{j},~M=\sum_{i}s_{i},
\end{equation}
where the spin variables $s_{i}=\pm1$ and the sum $\left\langle
ij\right\rangle $ considers nearest-neighbor interactions only. The
existence of a magnetic feedback effect $\left\langle\delta
H^{\omega}\delta M\right\rangle$ can be implemented using a
Metropolis algorithm \cite{met} with the acceptance probability:
\begin{equation}
p\left( U,M\left\vert U+\Delta U,M+\Delta M\right.
\right)=\min\left\{1, \exp\left[ -\beta\Delta U+\beta
H^{\omega}\Delta M\right] \right\}.  \label{as.met}
\end{equation}
Denoting by $m=M/N$ the magnetization per particle, the external
magnetic field in this study is given by
$H^{\omega}=\bar{H}+\lambda\left( m-\bar{m}\right) $, where $\bar{m}
$ and $\bar{H}$ are some roughly estimations of the expectation
values $ \left\langle m\right\rangle $ and $\left\langle
H^{\omega}\right\rangle $. Our goal is to obtain the isotherms of 2D
Ising model within anomalous regions with $\chi_{T}<0$. The
isothermal magnetic susceptibility per particle
$\bar{\chi}_{T}=\chi_{T}/N$ can be obtained from the fluctuation
relation (\ref{iso.sus2}) as follows:
\begin{equation}
\bar{\chi}_{T}^{-1}=\frac{1+\beta\left\langle \delta
H^{\omega}\delta
M\right\rangle }{\beta\left\langle \delta M^{2}\right\rangle }N\equiv \frac{%
1-\lambda\beta\sigma_{m}^{2}}{\beta\sigma_{m}^{2}},
\end{equation}
where $\sigma_{m}^{2}=\left\langle \delta M^{2}\right\rangle /N$
represents the thermal dispersion of magnetization. The values of
parameters $\left( \bar{H}, \bar{m}\right) $ can be provided using
the susceptibility per particle $\bar{\chi}_{T}$ obtained from a
previous MC calculation throughout the expression:
\begin{equation}
\bar{H}_{i+1}=\bar{H}_{i}+\left( \bar{\chi}_{T}\right) _{i}^{-1}\left( \bar{m%
}_{i+1}-\bar{m}_{i}\right) ,  \label{ser}
\end{equation}
where the step $\Delta m=\bar{m}_{i+1}-\bar{m}_{i}$ should be small.
Here, the initial value $\bar{m}_{0}$ is estimated as the average of
magnetization calculated from an ordinary Metropolis algorithm with
constant magnetic field $H^{\omega }=\bar{H}_{0}$ far enough from
the unstable region with $\chi_{T}<0$.
\begin{figure}[t]
\begin{center}
\includegraphics[
height=4.7in, width=3.4in ]{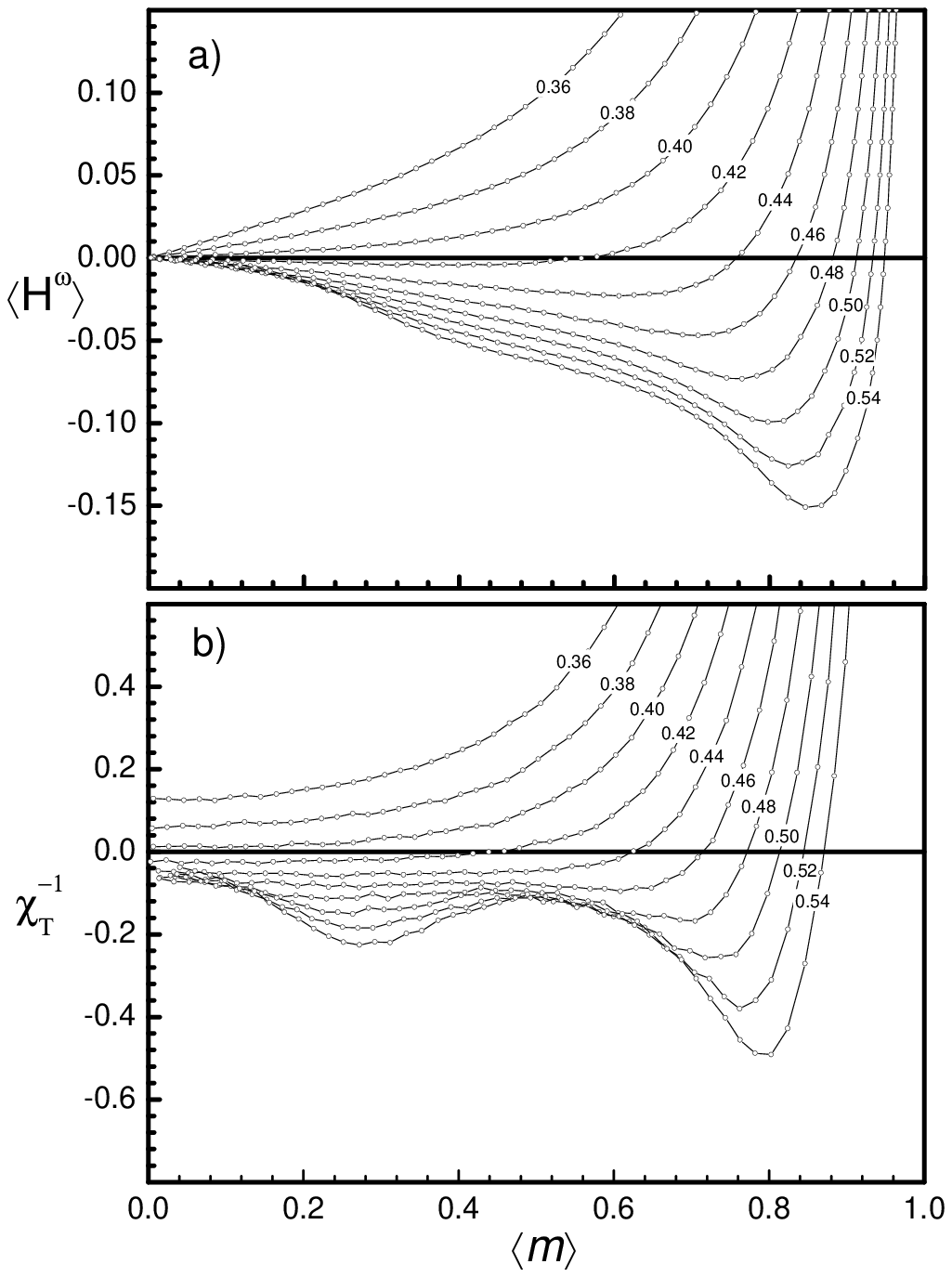}
\end{center}
\caption{Isotherms of 2D Ising model with $L=25$ derived from
Metropolis Monte Carlo simulations using the magnetic feedback
effect $\left\langle \delta H^{\omega}\delta M\right\rangle$: Panel
a) magnetic field $\left\langle H^{\omega }\right\rangle $
\textit{versus} magnetization $\left\langle m\right\rangle $ ; Panel
b) inverse isothermal magnetic susceptibility $\bar{\chi} _{T}^{-1}$
\textit{versus} magnetization $\left\langle m\right\rangle $. The
labels in both panels indicate the value of $\beta$ for the
corresponding isotherm.} \label{anomalous.eps}
\end{figure}
Although any real value of coupling constant $\lambda$ that
satisfies stability condition (\ref{mag.cond}) is admissible, one
can impose a constraint to reduce as low as possible the thermal
fluctuations of the system magnetization (\ref{iso.sus2}) and the
external magnetic field (\ref{iso.sus3}). According to these
expressions, the growth of the coupling constant $\lambda$ provokes
a reduction of the magnetization fluctuations and the growth of the
external magnetic field $H^{\omega}$ fluctuations. Due to this
observation, the optimal value of the coupling constant $\lambda$ is
chosen to minimize the total dispersion
$\sigma^{2}=\sigma_{H}^{2}+\sigma_{m}^{2}$, where:
\begin{equation}
\sigma_{H}^{2}=N\left\langle \left(\delta
H^{\omega}\right)^{2}\right\rangle
\end{equation}
is the thermal dispersion of the external magnetic field. Such an
analysis yields:
\begin{equation}
\lambda_{op}(a)=\sqrt{a^{2}+1}-a,
\end{equation}
where $a$ is the inverse of the isothermal magnetic susceptibility
per particle, $a=\bar{\chi}_{T}^{-1}$. Thus, the value of the
coupling constant employed in the $i+1$-calculation is estimated
from the previous MC calculation as:
\begin{equation}
\lambda_{i+1}=\lambda_{op}\left[(\bar{\chi}_{T})^{-1}_{i}\right].
\end{equation}

\begin{figure}[t]
\begin{center}
\includegraphics[
height=2.5in, width=3.4in ]{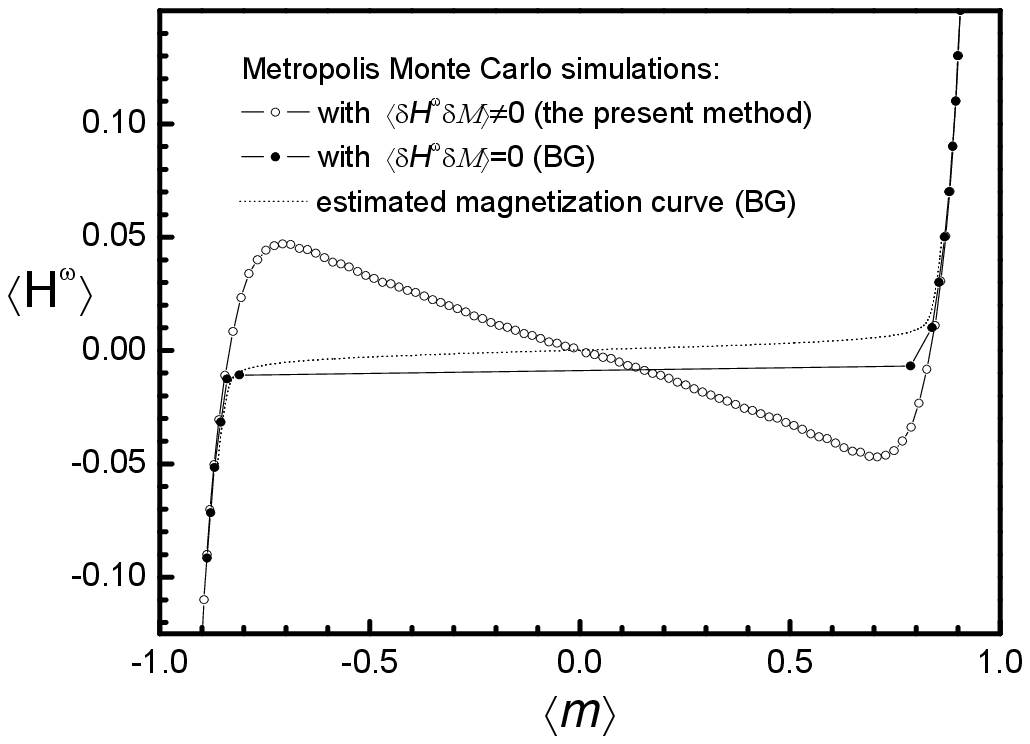}
\end{center}
\caption{Isotherms for $L=25$ and $\beta=0.46$ obtained from
Metropolis Monte Carlo simulations with (open circles) and without
(black circles) the incidence of a magnetic feedback effect
$\left\langle\delta H^{\omega}\delta M\right\rangle$. For comparison
purposes, it is also shown the magnetization curve $m(\beta,H)$
associated with the Boltzmann-Gibbs distributions (\ref{BGD})
obtained from the estimation procedure described in \ref{procedure}
(doted line).} \label{comparison.eps}
\end{figure}

Results of MC simulations using the procedure previously explained
are shown in FIG.\ref{anomalous.eps}. The simulations were
restricted to a lattice with $L=25$ and $n=10^{6}$ iterations for
each calculated point of these isotherms. Besides, states with
positive magnetization $\left\langle m\right\rangle>0$ are only
shown due to the existence of the symmetry $ M\rightarrow-M$ and
$H\rightarrow-H$. These results revealed the presence of anomalous
diamagnetic states $\bar{\chi}_{T}<0$ for inverse temperatures
$\beta$ above the critical point $\beta_{c}\simeq0.41$, that is, for
temperatures $T<T_{c}\simeq2.44$. Notice that these dependencies are
very similar to the ones shown in panels \textbf{a} and \textbf{b}
at FIG.\ref{Weiss.eps} corresponding to ferromagnetic Weiss model.

\begin{figure}[t]
\begin{center}
\includegraphics[
height=2.5in, width=3.4in ]{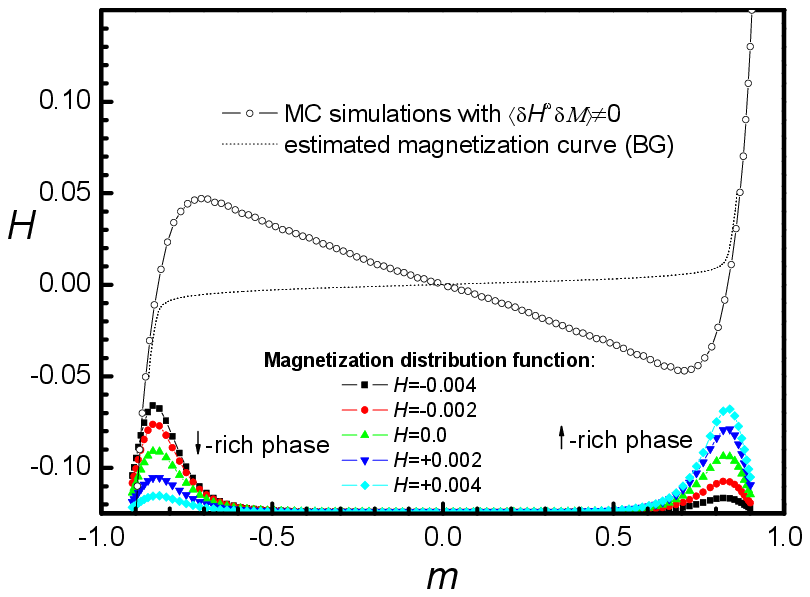}
\end{center}
\caption{Bimodal character of magnetization distribution function
$dq(M|\beta,H)$ for $L=25$, $\beta=0.46$ and small values of the
external magnetic field $H$ obtained from the procedure described in
\ref{procedure}. The coexisting peaks correspond to stable and
metastable configurations with positive isothermal magnetic
susceptibility $\chi_{T}>0$. Clearly, anomalous region with
diamagnetic states  $\chi_{T}<0$ is poorly accessed within the
Boltzmann-Gibbs distributions (\ref{BGD}), which excludes the
occurrence of a gradual transition from $\uparrow$-rich towards
$\downarrow$-rich domain configurations with the modification of the
external magnetic field.} \label{distributions.eps}
\end{figure}

We show in FIG.\ref{comparison.eps} a comparative study of three
different methods to obtain the isotherm with $L=25$ and
$\beta=0.46$ in the $H-M$ diagram: Metropolis MC simulations with
and without magnetic feedback effect, as well as the magnetization
curve $m(\beta,H)$ associated with Boltzmann-Gibbs distribution
(\ref{BGD}) derived from the estimation procedure described in
\ref{procedure}. While the existence of a magnetic feedback effect
$\left\langle\delta H^{\omega}\delta M\right\rangle$ allows to
reveal the backbending of this curve, its counterpart dependencies
at constant external magnetic field undergo a sudden change in
magnetization around the value $H=0$. Conventionally, such a sudden
change is interpreted as the occurrence of a discontinuous phase
transition. However, this behavior is just a consequence of the
inability of the environmental influence characterized by a constant
temperature and a constant external magnetic field to access the
region with anomalous response, which is clearly illustrated in
FIG.\ref{distributions.eps}.

Generally speaking, any isotherm obtained from Boltzmann-Gibbs
distributions (\ref{BGD}) is continuous and monotonous in the $H-M$
diagram for $L$ finite (as the doted line shown in
FIG.\ref{comparison.eps}), which converges towards the familiar
\textit{Maxwell construction} below critical temperature in the
thermodynamic limit $L\rightarrow\infty$. However, a direct
calculation of this curve is very difficult to carry out with a
sufficient precision using MC simulations at constant external
magnetic field. The discrepancy between the two BG dependencies
shown in FIG.\ref{comparison.eps} evidences a typical difficulty of
conventional MC simulations in presence of discontinuous phase
transitions: the incidence of a \textit{super-critical slowing down}
\cite{Binder}. This phenomenon manifests as a poor equilibration of
MC expectation values due to the exponential growth of convergence
times with the increase of the system size $L$. The origin of this
dynamical anomaly is the \textit{bimodal character} of the
magnetization distribution function $dq(M|\beta,H)$ below the
critical temperature and small values of $H$ (see in
FIG.\ref{distributions.eps}), which can induce an effective trapping
of MC dynamics in any of coexisting peaks. In particular, MC
simulations with constant magnetic field of FIG.\ref{comparison.eps}
undergo an effective trapping in the peak with positive
magnetization for small negative values of $H$.

Fortunately, the incidence of a magnetic feedback effect
$\left\langle\delta H^{\omega}\delta M\right\rangle$ can suppress
the thermodynamic instability associated with the discontinuous
phase transition of the 2D Ising model. Since the sudden jump does
not occur, one could claim that \textit{the discontinuous phase
transition observed within the framework of Boltzmann-Gibbs
distributions} (\ref{BGD}) \textit{has been suppressed in the
present environmental influence}. However, one observes the
coexistence of magnetic domains with different orientations during
the inversion of the system magnetization $M\rightarrow -M$. Along
this process, the transition from $\uparrow$-rich towards
$\downarrow$-rich domain configurations is gradual and without
metastability, which suggest that phase separation actually persist
at macroscopic level. Essentially, this type of behavior is
analogous to a gradual conversion of the liquid water towards its
solid phase without the incidence of metastability (as the trapping
of the system evolution in a metastable state as \textit{supercooled
liquid}).

\section{Concluding remarks}\label{final}
Usual equilibrium fluctuation theorems of statistical mechanics
disregard the existence of states with anomalous response. Starting
from general fluctuation theorems of any classical distribution
function (\ref{equi.cond}) and (\ref{fund.assoc}), we have been able
to obtain three equilibrium fluctuation theorems compatible with
anomalous response, Eqs.(\ref{fund.theorem})-(\ref{complementary}).
A novel feature is the consideration of \textit{environmental
feedback effects} described by the correlation matric $\mathcal{D}$,
which involves the existence of non-vanishing correlations among the
system observables $(U,X)$ and the environmental control variables
$(T^{\omega},Y^{\omega})$. As evidenced in the study of 2D Ising
model, these theorems can be successfully employed for the analysis
of the thermodynamic stability beyond the conventional equilibrium
situations of statistical mechanics.

\section*{Acknowledgements}

L Velazquez thanks the financial support of CONICYT/Programa
Bicentenario de Ciencia y Tecnolog\'{i}a PSD \textbf{65} (Chilean
agency).
\appendix
\section{Estimation procedure}\label{procedure}
Starting from Boltzmann-Gibbs distribution (\ref{BGD}), the
magnetization distribution function $ dq \left( M|\beta ,H\right) $
can be expressed as:
\begin{equation}
dq \left( M|\beta ,H\right) =\frac{1}{Z\left( \beta ,H\right) }\exp
\left( \beta HM\right) \Xi \left( M|\beta \right)dM ,
\end{equation}%
where the states density at constant temperature $\Xi \left( M|\beta
\right) $ is obtained from the states density $\Omega \left(
U,M\right) $:
\begin{equation}
\Xi \left( M|\beta \right) =\int \exp \left( -\beta U\right) \Omega
\left( U,M\right) dU,
\end{equation}%
while the partition function $Z\left( \beta ,H\right) $ by means of
the normalization condition:
\begin{equation}\label{normalization}
Z\left( \beta ,H\right) =\int \exp \left( \beta HM\right) \Xi \left(
M|\beta \right) dM.
\end{equation}%
Equilibrium thermodynamic conditions (\ref{eq}) allow to estimate
$\Xi \left( M|\beta \right) $ using the isotherm $\left\langle
H^{\omega}\right\rangle\left( M|\beta \right) $ obtained from MC
simulations with magnetic feedback effect:
\begin{equation}
\log \Xi \left( M|\beta \right) \simeq C-\beta
\int_{M_{1}}^{M}\left\langle H^{\omega}\right\rangle\left( M^{\prime
}|\beta \right) dM^{\prime },
\end{equation}%
where $C$ is specified with the normalization condition
(\ref{normalization}). The estimated magnetization distribution
function $dq\left( M|\beta ,H\right) $ can be employed to obtain the
magnetization curve $m(\beta,H)$ illustrated in
FIG.\ref{comparison.eps} and FIG.\ref{distributions.eps} as:
\begin{equation}
m(\beta,H) =\frac{1}{N}\int Mdq \left( M|\beta ,H\right),
\end{equation}
with $N$ being the system size.


\section*{References}

\begin{thebibliography}{99}
\bibitem{Lynden} Lynden-Bell D 1967 \textit{Mon. Not. R. Astro Soc.} \textbf{%
136}, 101; Lynden-Bell D and Wood R 1968 \textit{Mon. Not. R. Astro Soc.}
\textbf{138} 495.

\bibitem{moretto} Moretto L G, Ghetti R, Phair L, Tso K and Wozniak, G.J.
1997 \textit{Phys. Rep.} \textbf{287} 250.

\bibitem{Dagostino} D'Agostino \textit{et al} 2000 \textit{Phys. Lett.}
\textbf{B 473} 219.

\bibitem{gro na} Gross D H E and Madjet M E 1997 \textit{Z. Phys.} \textbf{B
104} 521.

\bibitem{Lyn2} Lynden-Bell D and Lynden-Bell R M 1977 \textit{Mon. Not. R.
Astro Soc.} \textbf{181} 405.

\bibitem{pad} Padmanabhan T 1990 \textit{Physics Reports} \textbf{188} 285.

\bibitem{Lyn3} Lynden-Bell D 1999 \textit{Physica A} \textbf{26} 293.

\bibitem{gro1} Gross D H E 2001 \textit{Microcanonical thermodynamics: Phase
transitions in Small systems}, \textit{66 Lectures Notes in Physics} (World
scientific, Singapore).

\bibitem{Dauxois} Dauxois T, Ruffo S, Arimondo E and Wilkens M (Eds.) 2002
\textit{Dynamics and Thermodynamics of Systems with Long Range Interactions}%
, \textit{Lecture Notes in Physics} (Springer, New York).

\bibitem{Vel.ETFR} Velazquez L and Curilef S 2009 \textit{J. Phys. A: Math.
Theor.} \textbf{42} 095006.

\bibitem{Vel.Thirring} Velazquez L and Curilef S 2009 \textit{J. Stat. Mech.: Theo.
Exp.} \textbf{P03027}.

\bibitem{Vel.URSM} Velazquez L and Curilef S 2009 \textit{Mod. Phys. Lett. B} \textbf{23
} 3551.

\bibitem{Vel.MC1} Velazquez L and Curilef S 2010 \textit{J. Stat. Mech.: Theo.
Exp.} \textbf{P02002}.

\bibitem{Vel.MC2} Velazquez L and Curilef S 2010 \textit{J. Stat. Mech.: Theo.
Exp.} \textbf{P04026}.

\bibitem{Vel.Geo} Velazquez L and Curilef S 2009 \textit{J. Phys. A: Math.
Theor.} \textbf{42} 335003.

\bibitem{Ison} Ison M J, Chernomoretz A and Dorso C O 2004 \textit{Physica}
\textbf{A 341} 389.

\bibitem{Einarsson} Einarsson B 2004 \textit{Phys. Lett.} \textbf{A 332} 335.

\bibitem{Chomaz} Chomaz P and Gulminelli F 2006 \textit{Eur. Phys. J.}
\textbf{A 30} 317.

\bibitem{Gulminelli} Gulminelli F 2007 \textit{Nuclear Physics} \textbf{A 791%
} 165.

\bibitem{Lovett} Lovett R 2007 \textit{Rep. Prog. Phys.} \textbf{70} 195.

\bibitem{Hugo} Campa A, Ruffo S and Touchette S 2007 \textit{Physica}
\textbf{A 385} 233.


\bibitem{landau} Landau L D and Lifzhitz E M 1977 \textit{Statistical Physics
} (Pergamon, New York).

\bibitem{Reichl} Reichl L E 1980 \textit{A modern course in Statistical
Mechanics}, (Univ. Texas Press, Austin).


\bibitem{Fisher} Fisher R A 1922 \textit{Philosophical Transactions, Royal Society of
London} (A), \textbf{222}, 309-368.

\bibitem{Rao} Rao C R 1945 \textit{Bull. Calcutta Math. Soc.} \textbf{37},
81-91.

\bibitem{Stanley} Stanley E 1971 \textit{Introduction to phase transitions and critical
phenomena}, (Claredon Press, Oxford).

\bibitem{met} Metropolis N, Rosenbluth A W, Rosenbluth M N, Teller A H and
Teller E 1953 \textit{J. Chem. Phys.} \textbf{21} 1087.

\bibitem{thir} Thirring W 1980 \textit{Quantum Mechanics of large systems}
(Springer) Ch. 2.3.

\bibitem{Binder} Landau P D and Binder K 2000 \textit{A Guide to Monte Carlo Simulations in
Statistical Physics} (Cambridge: Cambridge University Press).

\end{thebibliography}
\end{document}